\documentclass[preprint]{vgtc}               

\graphicspath{{figures/}{pictures/}{images/}{./}} 

\usepackage{times}                     

\usepackage{tabu}                      
\usepackage{booktabs}                  
\usepackage{lipsum}                    
\usepackage{mwe}                       
\usepackage{float}
\usepackage{tabularx}

\usepackage{mathptmx}                  
\usepackage{color,soul}
\usepackage{multirow}
\usepackage{arydshln} 
\newcommand{\vitem}{\vspace{-1.5mm}\item}
\usepackage{balance}

\usepackage{soul} 

\newif\ifhighlightchanges

\newcommand{\newmarker}[1]{%
\ifhighlightchanges
\textcolor{blue}{#1}%
\else
#1%
\fi
}

\highlightchangesfalse

\onlineid{1487}

\vgtccategory{Research}

\vgtcinsertpkg

\title{Long-Term Experiences From Working with Extended Reality in the Wild}

\author{Verena Biener\thanks{e-mail: verena.biener@visus.uni-stuttgart.de}
    \\ \scriptsize University of Stuttgart, Germany 
\and Florian Jack Winston
    \\ \scriptsize Graz University of Technology, Austria 
\and Dieter Schmalstieg
    \\ \scriptsize University of Stuttgart, Germany
\and Alexander Plopski\thanks{e-mail: alexander.plopski@tugraz.at}
    \\ \scriptsize Graz University of Technology, Austria
}

\teaser{
  \centering
  \includegraphics[width=\linewidth]{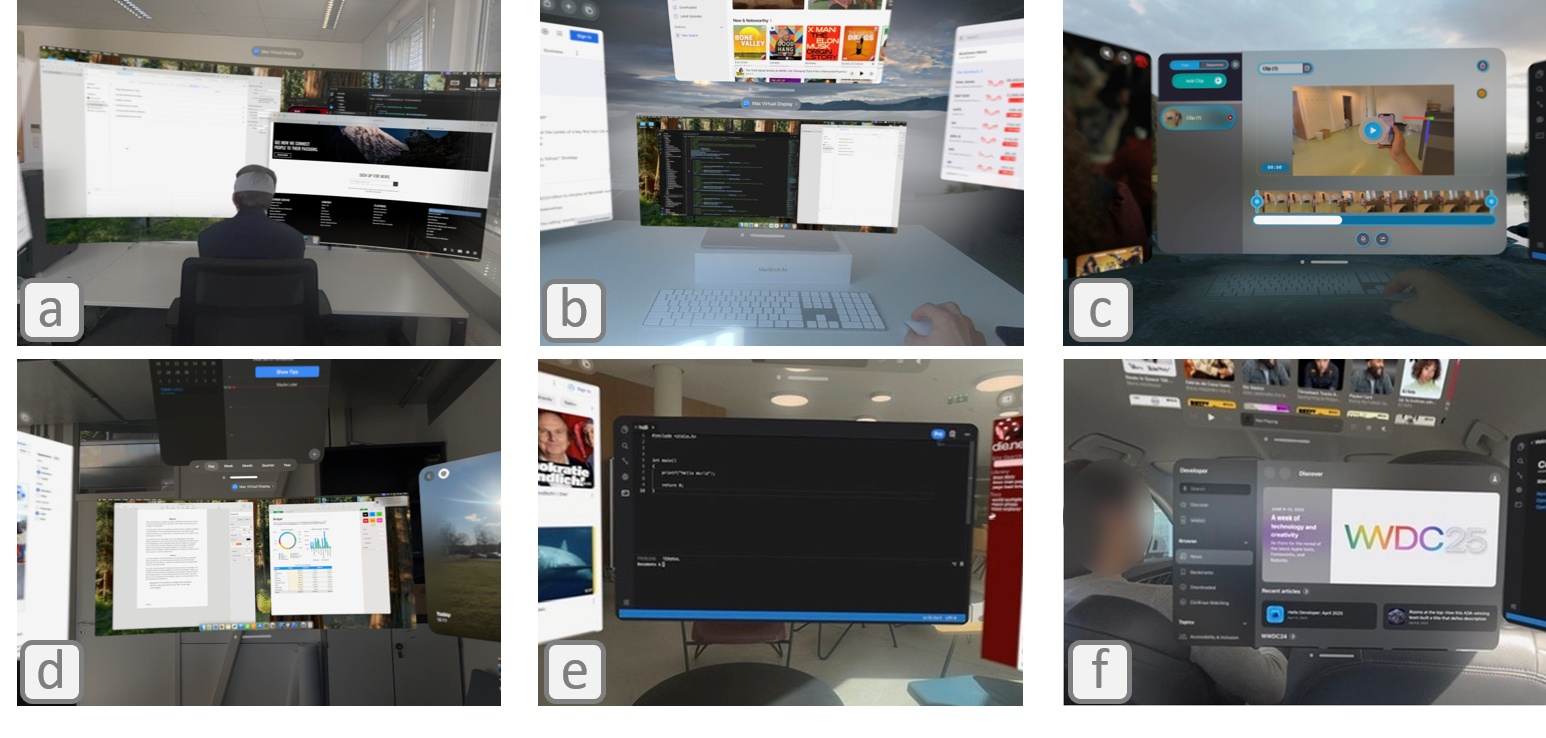}
  \caption{Top row: Different environments in which users work with the Apple Vision Pro: (a) Apple Vision Pro used as a wide field of view screen, (b) partial immersive environment, and (c) fully immersive environment (with hand and keyboard tracking). Bottom row: Different locations in which users use the Apple Vision Pro: (d) office, (e) public spaces, and (f) while traveling.} 
  \label{fig:teaser}
}

\abstract{
Extended Reality (XR) is increasingly used as a productivity tool and recent commercial XR devices have even been specifically designed as productivity tools, or, at least, are heavily advertised for such purposes, such as the Apple Vision Pro (AVP), which has now been available for more than one year.
In spite of what marketing suggests, research still lacks an understanding of the long-term usage of such devices in ecologically valid everyday settings, as most studies are conducted in very controlled environments.
Therefore, we conducted interviews with ten AVP users to better understand how experienced users engage with the device, and which limitations persist. 
Our participants report that XR can increase productivity and that they got used to the device after some time.
Yet, a range of limitations persist that might hinder the widespread use of XR as a productivity tool, such as a lack of native applications, difficulties when integrating XR into current workflows, and limited possibilities to adapt and customize the XR experience.
} 


\begin{document}


\firstsection{Introduction}

\maketitle


In recent years, much research has explored how Extended Reality (XR), which in this paper is meant to cover everything from augmented reality (AR) to virtual reality (VR), could be used to support knowledge or office workers.
Important themes in this research area cover the potentially large display space \cite{pavanatto2024multiple, lisle2023spaces, pavanatto2021we}, supporting specific tasks of knowledge workers \cite{biener2022povrpoint, gesslein2020pen, qiu2025marginalia,lu2023wild}, or reducing distractions \cite{ruvimova2020transport, lee2022partitioning}.
However, many studies rely on laboratory evaluations and lack genuine ecological validity. We also observe a lack of investigations on the long-term usage of XR devices for productivity. 
\newmarker{So far, the two longest studies lasted one full week \cite{biener2022quantifying} or were conducted over a period two weeks in which participants completed at least ten 30-minute sessions \cite{cheng2025augmented}.}
This paper aims to make a step towards closing the research gap on the real-life use of XR for knowledge work.
With recent commercial devices being promoted or even specifically designed as productivity tools, it becomes more feasible to study long-term effects. Such devices allow participants to use them for extended periods of time and meaningfully engage in real-world tasks or continue their personal daily tasks.
At the time of writing, the Apple Vision Pro (AVP) has been available for more than a year (released on February 2, 2024). In contrast to other XR headsets, which were primarily marketed for entertainment or training, the AVP is being advertised as a device for knowledge work with applications from the Apple Mac ecosystem. Although the device is expensive, it enjoys a dedicated community of users who bought an AVP for personal use. This situation gives us the opportunity to explore the usage of XR in a setting with superior ecological validity. 

Therefore, we conducted interviews with ten AVP users to gain a better understanding of their motivations for using the device. We wanted to learn how they are actually using it, what they value most about the device, what kind of workarounds they rely on, and what they desire most for future device iterations. Some examples of described use cases are shown in Figure~\ref{fig:teaser}.
All interviewed participants used the device regularly for at least six months and can provide insights from the point of view of experienced users. A comparable level of experience is not possible in a controlled experiment, where users might first need to learn how to use the device. We believe that the insight from this  ethnographic approach trumps any bias that may be introduced by the interviewees' status as early adopters or enthusiasts.
Our work's main contributions are:
\begin{itemize}
    \vitem We analyze publicly shared opinions of AVP users.
    \vitem We present interviews conducted with long-term users that use the device in a professional setting and demonstrate the potential of XR devices in this field.
    \vitem We discuss our findings in relation to prior findings and highlight remaining challenges from the perspective of users.
\end{itemize}


\section{Related Work}
\label{sec:related_work}

XR has gained popularity as a productivity tool, both in commercial products as well as in the research community. Here, we give a brief overview of recent developments using XR as a productivity tool and how XR has been evaluated in real-life scenarios so far. 

\subsection{XR as a Productivity Tool}

The potential of using XR in the work context has become popular in recent years. A wide range of research has evaluated various aspects of XR as a productivity tool~\cite{biener2023extended}.
XR has been shown to be valuable for tasks such as preparing presentation slides~\cite{biener2022povrpoint},  editing spreadsheets~\cite{gesslein2020pen}, unobtrusively acquiring information~\cite{lu2023wild}, or taking notes~\cite{qiu2025marginalia}.
One of the most prominent advantages of XR for knowledge work is its virtually unlimited screen real estate~\cite{Reichlen1993} that can increase productivity~\cite{czerwinski2003toward} and has uses as a  multi-window canvas, a cinematic screen, or a supplement to physical screens~\cite{pavanatto2024multiple}. 
A large display space has also been shown to increase user satisfaction and decrease frustration in a sense-making process~\cite{lisle2023spaces}.

Another advantage of using XR, as discussed in prior work, is the ability to reduce distractions through virtual content~\cite{ruvimova2020transport, lee2022partitioning} or to reduce stress~\cite{pretsch2020improving, thoondee2017using}. 
Using XR also increases privacy in public settings, when working with sensitive information~\cite{grubert2019office} or when typing passwords~\cite{schneider2019reconviguration}.
XR can also enable new modes of collaboration, such as telepresence~\cite{pejsa2016room2room}, collaborative code editing~\cite{dominic2020remote}, joint data analysis~\cite{butscher2018clusters, ens2020uplift}, or collaborative construction planning~\cite{dong2011collaborative}. 

Several commercial XR tools target knowledge work, including Varjo Workspace~\cite{varjoworkspace}, 
Immersed~\cite{immersed}, 
vSpatial~\cite{vspatial}, 
or Meta Horizon Workrooms~\cite{metahorizon}. 
Devices such as Lenovo Think Reality or Sightful's Spacetop are even specifically designed for knowledge work.
AVP is also marketed by focusing on its value as a work tool.  

\subsection{Limitations of XR}

Although XR can provide a wide range of advantages, current devices still have ergonomic limitations that can induce visual fatigue, muscle fatigue, acute stress, and mental overload~\cite{souchet2023narrative}.
The increased display space has its own set of limitations, such as a limited resolution or a greater need for head movements. Consequently, Pavanatto et al.~\cite{pavanatto2021we} recommend the use of a combination of physical and virtual monitors.
To handle increased head movements, McGill et al.~\cite{mcgill2020expanding} proposed to adjust the position of virtual displays depending on the user's gaze direction.

Other limitations emerge from the social context of using XR in public.
For example, prior work suggests that the placement of virtual content in public settings is not trivial and that users prefer to avoid collision of virtual content with physical objects or people~\cite{ng2021passenger, medeiros2022shielding, cheng2021semanticadapt}.
In some settings, it is very important to be aware of the physical surroundings. Therefore, McGill et al.~\cite{mcgill2015dose} proposed to integrate relevant parts of the physical surroundings into the virtual space. Several other authors have proposed techniques for doing so~\cite{wang2022realitylens, hartmann2019realitycheck, tao2022integrating, simeone2016vr, bajorunaite2023reality}.
One specific concern is the restricted view that XR devices allow on smartphones, which are considered essential devices used repeatedly throughout a workday. Therefore, previous work has investigated how smartphones can be accessed using immersive VR~\cite{alaee2018user, bai2021bringing, desai2017window}.
Keeping in touch with the physical surroundings is particularly important in public spaces when safety is at risk~\cite{bajorunaite2021virtual, eghbali2019social}. A recent study by Biener et al.~\cite{biener2024working} confirms that users feel safer if they can see the physical environment.

In addition to environmental awareness, social acceptability is a concern when using XR in the presence of other people.
Schwind et al.~\cite{schwind2018virtual} showed that social acceptability depends on the situation and is less acceptable in spaces that expect social interaction.
A major concern when using VR in public is the fear of being stared at while losing social awareness, as reported by Bajorunaite et al.~\cite{bajorunaite2021virtual}.
A similar concern is stated by Alallah et al.~\cite{alallah2018performer} who found that less noticeable input techniques are more socially acceptable.

Several of these limitations have been addressed in the design of the AVP, but also in other devices.
For example, physical keyboards can be rendered visible in the virtual environment, bystanders can be represented in the virtual world, and a smartphone screen can be streamed into the virtual space.

However, a major limitation of all this prior work is that most of the studies were conducted in a very controlled setting for rather short periods of time.
So far, only a limited number of studies have investigated the long-term usage of XR devices.
Among them are a 24-hour self-experiment~\cite{steinicke2014self} and studies in which participants spend 8 hours working in VR~\cite{guo2019mixed, guo2020exploring, shen2019mental}.
\newmarker{They found that compared to a standard office setup, a virtual office environment resulted in higher visual discomfort \cite{guo2020exploring} and higher mental fatigue \cite{shen2019mental}.}
In the longest study to date, 16 participants worked in VR for one week (8 hours per day)~\cite{biener2022quantifying} resulting in significantly worse ratings for measures such as simulator sickness or usability than in a non-VR workplace.
A follow-up article provided a more detailed analysis~\cite{biener2024hold} indicating that participants are gradually getting used to the HMD.
However, even these longitudinal studies were still conducted in a laboratory environment and have a finite duration of less than a week. 
Users of a complex XR system such as the AVP need time to explore all features and adjust their habits.

\subsection{Using XR In-The-Wild}

Several studies have explored the usage of XR in more ecologically valid settings. Lu et al.~\cite{lu2021evaluating, lu2023wild} conducted two in-the-wild studies in which participants used an AR prototype freely in everyday scenarios for several days.
To study the use of XR devices in public, Pavanatto et al.~\cite{pavanatto2024working} conducted two studies in which participants used XR devices in public for 30 minutes. 
In addition, Bajorunaite et al. \cite{bajorunaite2024vr} explored the use of XR on two 10-minute train rides, closely observed by the researchers who were conducting the study.
These studies all lasted for rather limited periods of time.
In contrast, Han et al. \cite{han2023people} reported on two studies in which, taken together, more than 200 university students used a VR device for weekly discussion sessions over the course of eight weeks.
\newmarker{A study by Tran et al. \cite{tran2025wearable} analyzed YouTube videos about experiences with different XR devices, but they did not focus on productivity work (only covered in 13\% of the analyzed videos). In addition, videos only provide restricted and very curated insights on user experiences.}
Cheng et al.~\cite{cheng2025augmented} conducted a diary study with 14 participants to investigate the usage of an AR laptop for daily work tasks. Within two weeks, the participants used the device at least ten times for a minimum of 30 minutes.
However, for longer periods, such a diary study could become too arduous for participants. Therefore, we decided to conduct detailed interviews with individuals who have extensively used an XR device (in our case, an AVP) for work.
Cheng et al. \cite{cheng2025augmented} suggest consulting power users as an alternative to time- and resource-consuming longitudinal studies.

\section{Methodology}
\label{sec:methodology}

The goal of our study was to explore how XR devices such as the AVP are used for longer periods of time in an ecologically valid setting.
We were interested whether this device actually works in users' daily lives and how they are using it, why they decided to use it, what they value most, and what the major drawbacks are.
To make such an investigation feasible within the time and budget constraints of academic research, we decided to find AVP users online who were already using the device for some time.
The advantage of our approach is that participants were using the device voluntarily and were not incentivized by any reward from a study.
We did not want to burden the participants with filling out questionnaires or keeping a diary over a long period of time. Therefore, we decided that interviewing users about their experiences was a considerate use of their time.
Even though an ethnographic approach can only give us qualitative data from a limited number of people and could be biased by individuals who favor the device, it represents real-life experiences and can challenge findings from lab-based research.

\newmarker{
In addition to the interviews, we performed a public sentiment analysis on the users of the r/AppleVisionPro and r/VisionPro subreddits. Posts and comments of both subreddits were examined between August and November of 2024 by sorting by "New" on both subreddits, examining each post individually and reading comments on posts related to work. Furthermore, posts between February and March 2024 were also examined. Only posts and comments of persons who worked for at least 40 hours with the AVP were included. This search resulted in 78 comments and posts with clear feedback relating to the AVP in a professional setting. They were sorted into one or more of 6 categories based on the most common themes within the posts and comments: screen, productivity/focus, ergonomics/comfort, travel, connectivity and miscellaneous. Each post and comment was then evaluated to be mostly positive or mostly negative.
}

\subsection{Positive Feedback in Public Sentiments}
The most common feedback, \newmarker{when evaluating the comments on reddit}, was the impact of the screen on the users. Five users reported that it was a great screen for work and five said that it was their main display for work. The ability to use multiple windows at once and spread them out through the AVP was useful for five users. One user commented on the sharpness of the display and how they did not suffer from eye fatigue. 
Nine users reported an increase in productivity. Three had an easier time focusing on their tasks and reducing the effects of hyperactivity disorder. It helped reduce anxiety for another user, and one user reported longer periods of uninterrupted work. Another used the AVP for utility apps while working on the Macbook. Two users liked the possibilities for spatial videos, while another called it a coding powerhouse.
Five users also noted the positive ergonomic effect, as the AVP allowed them to work from any position and location. It helped two of them alleviate back pain and thus increase productivity. One user said it is beneficial for them to get away from the desk, while it allowed another user to exercise while working. 
Three users find benefits of the AVP for travel, offering new opportunities for productivity and entertainment while traveling. \newmarker{They also reported, it} makes working with sensitive data more secure, as no outsider can see the screen.
Two users liked the connectivity of AVP, praising the device as great for remote work and extending the Apple ecosystem. One user called it incredible to work together remotely. 
Two users found the device easy to use and that it improved clarity and flexibility. Two users preferred spatial over flat computing, and one user said screen mirroring was smooth. One user called it the most intimate Apple device.

\subsection{Negative Feedback in Public Sentiments}
\newmarker{We also encountered negative feedback while evaluating comments on reddit.}
Two users complained that the ultra-wide display and low frame rate resulted in eye strain and motion sickness, and the text clarity was lower than 4k.
Two users found the AVP was not more productive than a high-performance workstation, but that it felt better and was more fun to use.
Another user said the AVP was less productive than a MacBook.
One user commented that it did not speed up any tasks, and two others could not see any use case for it. One user said putting on the device felt too much like a serious commitment, and another one called the device incredible for media consumption, but not for work. 
Compatibility to other Apple devices seemed to cause issues for three users, who noted difficulties connecting to the MacBook, keyboard, and mouse, or simply being unable to integrate the AVP into their setup. 
One user found the native apps to work better than the virtual display, while another said that many apps do not work well. 
Three users noted discomfort from heat and weight. One user needed to take breaks for balance and one mentioned a third-party strap to reduce neck pain. 
Two users found the AVP good but too expensive. One thought that the device was not yet ready and needed more improvements. Two users found that there were too many issues with the device.

\subsection{Creating Interview Questions}
We extended the insights from the public sentiment analysis with the findings and open questions mentioned in relevant research (\ref{sec:related_work}) to create a set of questions that were both interesting to our research and important to the users, according to the public posts.
The questions can be roughly grouped into demographics, purchase decision, usage patterns, and feature requests.
A full list of questions can be found in the supplementary material.

\subsection{Participants Recruitment}

We reached out to active users identified during our topic review with recruitment messages on subreddits r/AppleVisionPro and r/VisionPro, the Discord server associated with r/VisionPro, and we organized an event on the inSpaze~\cite{inspaze} platform, \newmarker{a social application for AVP users}. The posts were approved by the moderators of the corresponding messaging boards. To participate in the study, participants needed at least 40 hours of AVP use in a professional capacity.
These advertisements resulted in ten interviews. Participants registered on the Prolific platform to take part in our registered study. They signed an informed consent form, followed by selecting a time for the interview. After the interview, the participants were paid around 15 EUR through the prolific platform. Four participants agreed to be interviewed but refused remunerations through Prolific.

\subsection{Interview Procedure}

The participants were individually interviewed through WebEx, and the sessions were recorded.  
We started with questions about their person and profession and continued with open-ended questions. The questions addressed device expectations, experiences, advantages, drawbacks, and details on device usage. \newmarker{See the supplementary material for a list of all questions.}
There was no time limit on the responses, giving participants the option to give their full perspective. 
The predefined questions were used as a guide, but the interviewer adapted them dynamically. For example, we skipped questions that had already been answered or added follow-up questions depending on what the participant said.
Finally, they were asked whether there was anything to add or a topic was overlooked. 
The interviews typically lasted 30-60 minutes and consisted of 44 questions. The procedure was approved by the institutional review board of the University of Stuttgart.

\subsection{Analyzing the Interviews}
\newmarker{
Interviews were recorded during the session and later transcribed locally, using "OpenAI Whisper"~\cite{whisper} to preserve data privacy. Then we applied thematic analysis to investigate the data.
First, for each individual question, we extracted and summarized each participant's answer to ensure that all perspectives were included in our data.
To reduce redundancy, some statements were matched with other questions, in case participants drifted from the original question. In this process, some of the original questions were also merged into one topic if we found that answers were overlapping or addressing the same underlying theme.
Once organized, we compared the statements between participants to identify patterns and we merged similar answers to present them concisely. 
}

\section{Results}
Guided by the interview questions and through recurring themes during the interviews, we structured the results into 11 topics that are presented in sections \ref{subsec:incentive} to \ref{subsec:systemlimitations}. 

\subsection{Participants}

In total, ten AVP users participated in the interviews.
Their mean age was 48.8 years ($sd=14.68$), with the youngest person being 22, and the oldest, 68.
Nine participants were male, and one was female.
Six participants live in the US, two in Canada, one in the UK, and one in Belgium.
Four participants were self-employed.
The  participants' areas of work were diverse, ranging from video editing, biophysics, law, software and IT, consulting, public affairs, fashion, to journalism, without any notable groups.

On a scale from 1 (not at all) to 5 (extremely), participants rated their familiarity with Apple's ecosystem as 4.95 ($sd=0.16$), their familiarity with augmented reality as 4.0 ($sd=0.97$), and their familiarity with virtual reality as 3.75 ($sd=1.09$).

We started by asking participants about their history of using XR before the AVP.
Four participants (P1, P6, P7, P10) had owned another XR device. P5 was familiar with other headsets. P2 and P9 had briefly used another device.
Three participants (P3, P4, P5) had used hand-held AR or Google Cardboard before, while P8 had not used any other device before.
Several participants (P1, P3, P7) had used XR primarily for gaming before AVP, but P7 had also used the Meta Quest 2 for productivity tasks before.
P1 and P7 see the AVP as an XR device that can be used to perform serious knowledge work.
P10 initially used VR (Quest) in the context of a museum, and P1 had used the HTC Vive at a start-up several years ago but found the technology not good enough back then.
P3 reported that his interest in XR was triggered by science fiction at a young age.

Six participants had picked up their AVP in February 2024, most of them on launch day.
One participant has it since April; two, since July (Canadian launch), and another one, since October.
All of them are still using the device regularly.
Although we had not asked them to do so, three participants (P1, P5, P9) joined the interview using their AVP persona, \newmarker{which is a 3D representation of the users, created by capturing images of their facial expressions.}

\subsection{Incentive for Using AVP}
\label{subsec:incentive}

Concerning the initial reasons for getting an AVP, two participants (P3, P7) had wanted to use the AVP specifically for work. While P3 wanted to eliminate all physical monitors, P7 intended to replace an iPad. P2 and P3 liked the idea of working with a larger-than-usual screen, and P3 wanted to increase his efficiency, which also included fading out the physical environment. He mentions that AVP provides ``just so many capabilities that you can't really get with any other platform at all''. Others (P2, P3, P7) decided to get the device out of general interest or a passion for XR. Another group of participants (P4, P7, P8, P9) wanted to test what the device is capable of; P7 specifically mentioned hand and eye tracking. Participants also decided to buy the AVP because they felt a hype, and it seemed like a good time (P2, P6, P9, P10), or because they are long-term Apple users (P1, P2, P3, P7). P1 and P5 were initially introduced to the device through their work, and P10 was interested in using it for her VR museum. Several participants (P2, P3, P9) expected it to be good for entertainment. P8 specifically decided to get the AVP because he does not like to look down at a phone.


\begin{figure}[t]
	\centering
	\includegraphics[width=0.93\linewidth]{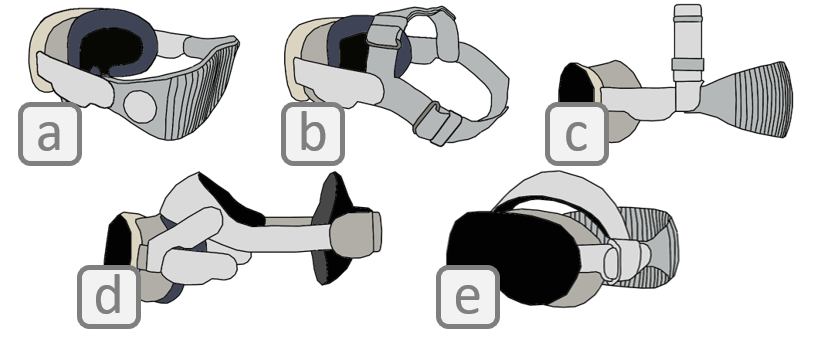}
	\caption{The head-straps that the participants from our interview-study are currently using: a) the Solo Knit Band, b) the Dual Loop Band, c) Belkin Head Strap, d) Globular Cluster, e) Anapro.    }
	\label{fig:straps}
    \vspace{-0.4cm}
\end{figure}
\subsection{Changes Over Time}

P4 and P7 are now using the device for much longer periods of time than they initially thought.
For P6 and P9, it is their primary device for both work and entertainment.
In contrast, P3 barely uses it outside of work, a choice he did not expect at the beginning.
P1 forgets about the narrow field of view unless one specifically focuses on it.
Six participants (P1, P2, P3, P4, P5, P8, P10) highlighted that the software has improved significantly since the launch of the AVP, especially with regard to the larger virtual display with a higher resolution, the improved persona and the gesture interaction.
However, P5 mentions that the device turned out more like an iPad than a Mac. P7 says that the virtual display is crucial for productivity.
P6 expressed some disappointment, saying that ``I was expecting a bigger increase of applications in a quicker way''.

\subsection{Usage Frequency}

All participants still use the AVP regularly.
The usage of some participants (P1, P2, P4, P6, P10) varies, but they use it several days a week, often between one and three hours.
P3 uses it three days a week for eight hours each when in the office.
The rest of the participants (P5, P7, P8, P9) use the device daily, at least five hours, some up to eight hours.
P5 and P8 take short breaks, but P5 says his behavior is comparable to that in a conventional office setup.

\subsection{Physical Comfort}

Five participants were using the standard head straps: Four (P2, P5, P7, P9) use the Solo knit band, and one (P4) uses the dual loop band. P7 uses a third-party head strap when traveling. Four participants (P1, P6, P8, P10) chose third-party straps (Belkin~\cite{belkin}, Globular Cluster~\cite{globular}, and Anapro~\cite{anapro}), while P3 built his own custom solution.
These head-straps are shown in Figure \ref{fig:straps}.
Most of the participants tried several different head straps before finding the most comfortable one for themselves. These choices appear to be highly individual and subjective. 
P1 describes it as ``virtual reality is not a one-size-fits-all experience''.
Three participants (P1, P3, P6) like to remove the light seal (the frame around the display that blocks the peripheral view of the real world) to have the impression of an increased field of view when in AR mode.

Six participants (P1, P2, P3 P4, P7, P9) grew accustomed to the weight and discomfort.
P3 and P8 continue to use the AVP even if it feels uncomfortable; P10 sometimes feels a little queasy in the morning, but not enough to make her stop using it. For P3, it is easier to blend out these issues when working on a work task than when using the headset for entertainment.
In contrast, P6 cannot wear the device for more than an hour without getting a headache.

\begin{figure}[t]
	\centering
	\includegraphics[width=\linewidth]{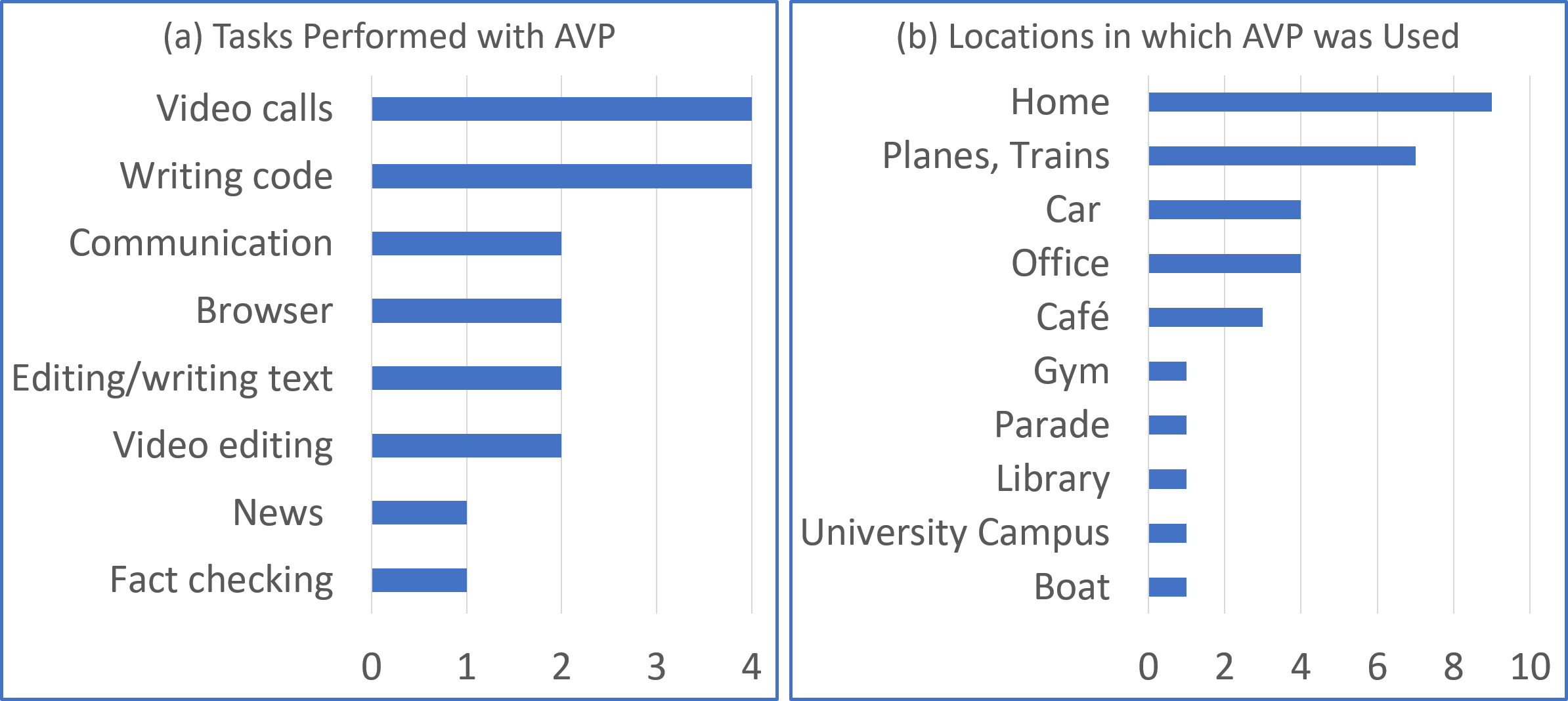}
	\caption{\newmarker{Number of participants that mentioned these tasks and locations during the interviews.}
        }
	\label{fig:summary}
    \vspace{-0.4cm}
\end{figure}

\subsection{Tasks Performed with AVP}

\begin{figure}[t]
	\centering
	\includegraphics[width=\linewidth]{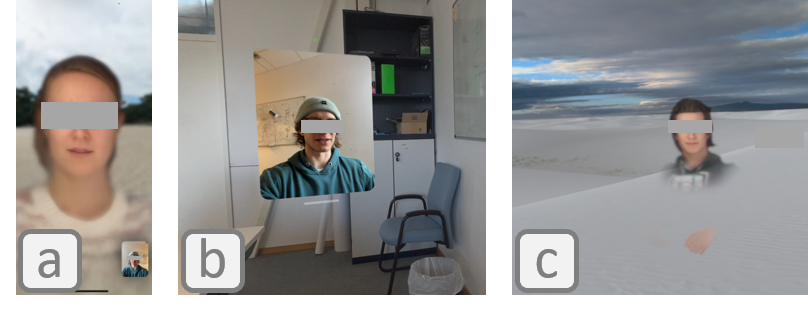}
	\caption{Video calls with AVP users as perceived by non-AVP users joining the call with a phone (a). The view of an AVP user when talking to a non-AVP user (b) or another AVP user (c).
        }
	\label{fig:videocalls}
    \vspace{-0.4cm}
\end{figure}

Four participants (P1, P2, P4, P6) use the AVP for parts of their work.
Others (P3, P5, P7, P8, P9) use it for most or all of their work.
\newmarker{Participants perform different tasks with the AVP which are listed in Fig. \ref{fig:summary}a}, and some use-cases are visible in Figure~\ref{fig:teaser}.
Using AVP for video calls was slightly controversial. P1 would only engage in video calls if all attendees are using a headset, and P2 is usually turning off the camera. P6 is not using AVP for video calls, because colleagues found that the avatar was too freakish.
P9 stated that it feels ``asymmetric'' if a previously recorded persona talks to a person in a live video.
Different scenarios of using the AVP for video calls are shown in Figure \ref{fig:videocalls}.
All except two (P3, P5) use the device for entertainment.
P10 was the only participant who does not performing serious knowledge work with the AVP. Besides entertainment, she used it to demo the applications her company is developing for the AVP and to write her diary.

\subsection{Usage Scenarios}

\newmarker{Participants used the AVP in a range of private and public locations. We list them in Fig. \ref{fig:summary}b and depict some of these in Figure~\ref{fig:teaser}.}
P8 goes nowhere without the device, and P7 used it anywhere if he has some downtime.
P7 uses it even while eating and drinking, albeit not with his family.
Eight out of ten participants (P1, P3, P4, P5, P7, P8, P9, P10) have used the AVP in public.
They reported only a few issues when using the device in public.
These include missing something important from the physical surroundings that can be solved using the pass-through view in critical directions (P1). P1 mentioned that not all places feel safe enough, but planes and trains are acceptable.
Some frequently encounter bystanders who are looking or commenting (P1, P4, P5), but P3 found that curiosity wears off quickly. P8 told us that his family members feel strange about him using the device in public.
P2 is not brave enough to use it in public until XR devices are more widespread, stating ``I’m not going to be the first guy''. 
P1 believes that there are more social challenges than technical ones. 
One challenge many XR headsets face is the occlusion of the user's face by the device. P7 considers the eyesight feature that shows the user's eyes on the outside of the display to be the primary reason his family accepted him using the device extensively at home. By seeing the eyes shown on the outside, they felt less disconnected. P8 is not using that feature, probably due to privacy concerns, although this behavior sometimes causes arguments with his wife.
P9 experienced that bystanders find the eyesight feature rather unnatural. 

\subsection{Productivity}

All except two (P4, P10) find the AVP makes them more productive.
The main reasons for the increased productivity are the ample display space that allows the user to see everything at once without the need to switch windows (P1, P2, P7) and that the AVP suppresses distractions and increases flow (P2, P5, P9).
P8 mentions that text can be displayed in a larger and better readable manner. P5 believes ergonomics are better than with desktop computers, and P7 states that the mobility afforded by the AVP increases productivity.
P3 even finds he is ``so productive with it that I end up having more free time''. P8 says, ``It changed my life. It extended my career''.
P4 feels his productivity is about the same as before, and P2 specifies that the productivity is approximately the same as when working with a large physical monitor.
However, P9 admits that sometimes he does not get anything done, even with many windows open.
P10 thinks she will be more productive in the AVP in the near future and will do more actual work in it when the interfaces are perfected.

Most of the participants (P2, P3, P4, P5, P7, P8, P10) mention that the ability to suppress the real world and be only surrounded by work helps them to better concentrate.
P3 describes that the ritual of turning on the device puts him in work mode.
P9 mentions that AVP does not affect his ability to concentrate on work differently than other technologies, while P6 found it even harder to concentrate with AVP, as he never forgets that he is wearing the device.
Participants also reported new types of distractions (P1, P7), such as new apps.


\subsection{Setup}

All except P10 are using both native apps and screen mirroring from their Mac.
P10 is only using native apps so far.
P2 and P3 prefer mirroring for some tasks, because the Mac apps that are already set up provide continuity.
All participants describe their virtual setup in a strikingly similar way.
They have their main window or application in front; often this front view shows the mirrored Mac screen.
On the sides, they are placing applications such as music (P1, P6), notes (P1, P8), communication and collaboration apps (P2, P3, P9), a web browser (P2, P7), calendar (P3, P5), mail (P6, P7), social media (P7), videos or podcasts (P7, P9, P10), photos (P10), a menu bar (P9), or the debug view of a currently developed app (P4). 
For all these purposes, participants often use native AVP apps.
Figure~\ref{fig:teaser} shows representative images of such setups.
Some participants (P2, P3, P5) put content above their heads to keep it glanceable, but not in focus. An apparent reason is the lack of a minimize operation.

Regarding the physical setup, all participants use a physical keyboard, either the one of their MacBook, or a dedicated one.
Furthermore, all participants, except P10, reported using a mouse (P2, P3, P6, P7, P8) or a trackpad (P1, P3, P4, P5, P7, P9).
P1 uses an iPad with a pencil for illustrations.
Apple keyboards can be detected and displayed in immersive mode. However, since these keyboards did not meet the requirements of P8, he added Velcro tape to help place his fingers in the correct position.

We also asked the participants if the AVP replaced any of their existing devices.
Several participants completely replaced their iPad (P2, P7) or use it much less (P1, P5).
Some participants replaced their monitors to varying degrees (P3, P4, P6, P8, P9). For example, P3 no longer has any monitors in the office, and P4 does not need to bring an external display when working on his boat.
P3 reports that the AVP has also partially replaced his laptop when going to the office and P10 is using her laptop less, while for P5 it replaced his Mac for many tasks.
P8 and P9 use their smartphones less, for P8 it replaces the TV when watching alone, and it fully replaced the TV for P10 who is now watching much more.

\subsection{Immersiveness}

All participants were using both immersive environments and the passthrough mode to some degree.
They reported that they prefer passthrough whenever they need to be aware of their surroundings (P1, P5), when there are other people around (P1, P2, P7, P8), while traveling (P5), when just browsing (P2), or when they are at home without distractions (P9).
Reasons for using the immersive environment are to eliminate distractions (P1, P2, P4, P5 P8), to work (P3, P8), or to relax (P9). 
They also use the immersive environment when watching movies (P1, P3, P6, P10) and when traveling (P9).
P4 specifically mentioned that he uses the immersive environment in coffee shops to shut out the physical world and that approaching people are not a concern.
P5 mentions that he used immersive environments a lot in the beginning, but that the novelty wore off, and P10 is not so impressed by background choices and would prefer a totally black one.
All but one participant (P3) report that they do not feel isolated while using AVP. 
P3 feels isolated, but states that it is not a problem for him. 
P4 generally does not feel isolated, but mentions that he sometimes intentionally isolates himself. 

\subsection{Advantages}

When asked about the main advantages of the AVP, most of the participants (P1, P2, P3, P4, P5, P6, P7, P9) highlight the large screen space.
P1 and P5 mention the ability to redesign the workspace, making work flexible but organized (P2).
P2 also highlights that the AVP provides a distraction-free environment.
Some participants (P3, P7, P8, P9) like portability; P3 and P7 specifically praised the ability to walk around and do other things while using the device.
A related advantage mentioned by P7 is the option of hands-free operation.
P9 thinks that a laptop in combination with AVP is equivalent to a desktop setup, and P1 specifically likes the integration of the AVP with existing (Apple) products.
One participant (P4) mentions stereo vision as an advantage, while P7, P9, and P10 think it is the best device for entertainment.
P10 also highly values spatial photography, specifically for watching her grandchildren who live abroad.
P5 likes the persona feature, which he always uses for video calls because the background or clothes no longer matter.

\subsection{System Limitations}
\label{subsec:systemlimitations}

Unsurprisingly, several participants criticized the weight and comfort of the device (P1, P3, P8, P9, P10).
In addition, P4 and P9 mentioned that the passthrough is not yet good enough, especially when trying to read something in the physical world, and P6 does not like the motion blur.
P1 does not like the limited field of view and prefers to use the device without the light seal.
Two participants (P6, P9) think that there are not enough native apps, and P1 reports accuracy issues when interacting with certain apps that were originally developed for the iPad.
P3 mentions that many collaborative experiences are limited because not many people own an AVP and P1 reports that collaboration through SharePlay is still limited to certain apps.
Other things that the participants criticized concern the limited battery life (P1, P10), missing haptic feedback (P1), the clunky virtual keyboard (P1), cable management (P4), the limitation of the people awareness feature (P7), and missing AI capabilities (P1).
As a developer, P5 complains that it is not possible to use the AVP for AVP development.
Some participants mention usage restrictions: for example, the device is not waterproof (P7) and still too novel to use in some demanding situations (P4).
Finally, P8 is concerned about the long-term effects of using such devices.

We asked the participants which features they desire most.
The participants would like a lighter device (P2, P3, P6), longer battery life (P2), and a wider field of view (P3). P3 also suggests a brighter display, and P7 would like to have faster foveated rendering.
In terms of hardware, P6 and P7 would like to have a controller, and P8 would like to have reduced sound spillage.
Concerning the software, participants would like to have multiple screens (P3, P9). P3 would like to be able to use touch interaction on the mirrored screen and P10 would like more sensitive interaction possibilities such as a pen for drawing.
The participants would like easier device sharing (P4, P9); P9 would even like to remote-control AVP in guest mode. Several participants would like to see more apps (P6, P7, P9); P5 wanted the ability to run MacOS natively.
P4 and P5 would like to grant camera access to third-party apps for a better experience.
P1 wishes for improved collaboration possibilities, both for people in the same room and in remote physical locations.
P9 suggested automatically unlocking nearby devices, being able to include the smartphone screen and phone-call features into AVP, pinning a window so it follows the user around, a save feature for workspaces, haptics through a pencil or smartwatch, multiple personas for different situations, customization of the content shown on the eyesight (exterior) display, and the ability to disable collision warnings.

Activities that participants would not want to do with AVP include exercising (P2, P8, P9), \newmarker{mainly because of the increased sweating}, driving (P3, P4, P5, P8), going for a walk (P3, P7, P10), anything not socially acceptable \newmarker{due to the built-in camera} (P7), or invasions of the space of other people (P10).

Several participants imagine that, in 5-10 years, many people will use devices like the AVP, and are excited to see how it evolves.



\section{Discussion}
\newmarker{This section provides a list of design recommendations that we derived from the results of our interview study. In addition, we further compare our findings to those of prior research. An overview of that comparison is shown in Table \ref{tab:comparison-to-prior-work}.}

\begin{table*}[h]
  \centering
    \caption{Comparison of interview findings with prior research.}
    \renewcommand{\arraystretch}{1.5}
    \begin{tabular}{|p{.99\columnwidth} |p{.99\columnwidth}| } 
    \hline
    \textbf{Findings from Interviews} &\textbf{Prior Research} \\
    \hline
    Main area of work is usually in front of the user with secondary apps in the periphery. & \textbf{Matches} findings of prior studies \cite{cheng2025augmented,pavanatto2024working, lischke2016screen} \\
    \hdashline 
    Not all places are perceived as equally safe. & \textbf{Matches} findings by Pavanatto et al. \cite{pavanatto2024working}  \\ 
    \hdashline
    Awareness of the physical environment is important. & \textbf{Matches} findings of prior studies \cite{mcgill2015dose, pavanatto2024working, bajorunaite2024vr} \\
    \hdashline
    \newmarker{Not all users feel comfortable using XR in public.} & \newmarker{\textbf{Matches} findings of prior studies \cite{cheng2025augmented, pavanatto2024working}}\\
    \hdashline
    \newmarker{Weight and comfort of head-mounted displays is still an issue.} & \newmarker{\textbf{Matches} findings of prior studies \cite{biener2022quantifying, biener2024hold, souchet2023narrative, cheng2025augmented}} \\
    \hdashline
    Increased display space is one of the major advantages. & \textbf{Matches} findings of prior studies \cite{pavanatto2021we, pavanatto2024multiple, lisle2023spaces} \\ 
    \hdashline
    XR allows higher productivity through reduced distractions. & \textbf{Matches} findings of prior studies \cite{ruvimova2020transport, lee2022partitioning, wienrich2025immersion} \\
    \hdashline
    Placement of virtual content was not influenced by bystanders. & \textbf{Contradicts} findings of prior studies~\cite{ng2021passenger, medeiros2022shielding} \\
    \hdashline
    Users seem to get used to the device despite it’s drawbacks. & \textbf{Matches} findings by Biener et al. \cite{biener2022quantifying}\\
    \hdashline
    Benefits of XR can already outweigh the drawbacks & \textbf{Contradicts} Biener et al. \cite{biener2022quantifying}\\
    \hline
    \end{tabular}
    \vspace{-0.4cm}
  \label{tab:comparison-to-prior-work}
\end{table*}

\subsection{Design Considerations}
\newmarker{
{\textbf{XR should be easily integrable into existing workflows.}}
For several participants, the reason to test the AVP was its integration into a familiar ecosystem. In particular, all participants were very familiar with the Apple ecosystem. 
They also switched between headset and desktop systems or used them in combination with their existing infrastructure.
The mirrored screen played an important role as it allowed them to seamlessly continue in their workflow.
The public sentiment showed that compatibility was an important issue for multiple users who mentioned difficulties getting the AVP integrated into their setup.
Therefore, integration of future devices into existing applications and architectures will play a vital role and integration of new devices into current workflows should be made easy.
}

\newmarker{
{\textbf{The availability of native apps is crucial.}}
Several participants mentioned a lack of native applications.
The availability of apps will be crucial for long-term success, as it makes the system more independent from a desktop setup and, therefore, more flexible, such as when using it in mobile settings.
However, apps might not be easily transferrable from iPad or Mac to the XR system, as indicated by some usability problems with iPad apps on the AVP.
}

\newmarker{
{\textbf{Some apps need to be specifically adapted for XR.}}
Apps that were developed for the iPad can also be used on the AVP. While this provides a large amount of apps that can be used on the AVP, our findings suggest that not all of them might be well suited for use on the AVP. Therefore, developers should consider making some changes to the design or interface of their iPad apps, specifically considering input modalities such as gaze and gestures, before deploying them to the AVP. Tran et al.~\cite{tran2025wearable} also concluded that apps need to be re-imagined to exploit the full potential of XR.
}

\newmarker{
{\textbf{XR should support collaboration and device sharing.}}
Participants acknowledge the value of XR for collaborations, but in its current state it seems not to be fully developed. Some participants criticize limited possibilities to collaborate or share experiences with co-located or remote users.
Video calls, using the virtual persona, get mixed reviews. While it provides the advantage of appearing professional in every situation and location, it seems less appropriate when communicating with people who are not using a persona themselves, as it could make the conversation asymmetric.
Participants also expressed the desire to have multiple personas for different social situations, which could easily be implemented.
This feature would also support sharing of a device, as multiple participants criticized sharing as currently very inconvenient.
The eyesight feature was introduced in the AVP to support interaction between the XR user and other people. It seems to work well for some, while others do not use it at all. 
One participant also expressed the desire to expand this feature and make it customizable to convey more information to bystanders.
}

\newmarker{
{\textbf{XR should allow creating multiple virtual screens.}}
Most participants reported distributing applications into a main work area in front of them and secondary applications in the periphery. This arrangement is similar to prior studies~\cite{cheng2025augmented,pavanatto2024working, lischke2016screen} who looked at how XR is used in-the-wild (Table \ref{tab:comparison-to-prior-work}). However, when mirroring the computer monitor, the AVP allows only a single screen to be placed, which was criticized by some participants. 
It limits the possibilities to arrange the virtual setup, as additional screens can only be added through native applications, and some tools are currently only available through mirroring a Mac. 
However, multiple virtual screens might not be necessary, as Pavanatto et al. \cite{pavanatto2024multiple} found no clear evidence that a single canvas or multiple screens provide a better user experience. However, the size of the canvas in their study spanned the same space as multiple displays would, which is not currently true for the AVP.
}

\newmarker{
{\textbf{XR should support place, resize and minimize.}}
While participants value the flexibility of the virtual environment that allows them to design their own virtual workplace, they are missing certain functionalities such as saving different setups for certain locations or tasks, or attaching windows to the display to follow the users as they walk around.
Current research is already looking into this, for example, how virtual content can be arranged in space by using gestures \cite{luo2025documents}, how content can semantically be adapted to the physical environment \cite{cheng2021semanticadapt},  how content can be placed according to user-driven constraints \cite{niyazov2023user}, how the optimal content placement can be computed by analyzing user behavior \cite{fender2017heatspace}, how content could transition between different positions in an automatic way \cite{ dhaka2024exploring}, or how virtual content can adapt to user movement \cite{lages2019walking}.
These approaches are highly relevant, and promising techniques should be incorporated into current XR devices.
Participants also mentioned they would like more AI capabilities. AI could play an important role in supporting the user with setting up their environment, for example, by using reinforcement learning to optimize automatic content placement~\cite{lu2024adaptive}, or by using a vision-language model to adjust virtual content to environmental or social cues \cite{li2024situationadapt}.
}

\newmarker{
Participants also mentioned that the system lacks a minimizing function for applications, so they often place them outside their field of view, such as above their heads. Future systems and applications could give users more flexibility to resize or place content. App windows could be resizable into a different format. Resizing might limit the available functionality, but could make it possible to place them close to the main screen without taking up much space. This could be combined with adaptive resizing accounting for the user's task, focus, and mental workload~\cite{lindlbauer2019context}.
}

\newmarker{
{\textbf{XR should provide physical environment awareness.}}
All but one participant acknowledge that awareness of the physical surroundings in public spaces is important and they would use some degree of passthrough in such situations. These statements match the findings of previous research that a certain degree of awareness of the physical world is important \cite{mcgill2015dose, pavanatto2024working, bajorunaite2024vr} (Table \ref{tab:comparison-to-prior-work}).
It was also mentioned that the current feature of detecting and indicating bystanders ("people awareness") does not yet work well enough. Researchers are aware of the problem of integrating bystanders into the virtual environment and have already proposed some solutions, such as using physical objects as anchors to show cues from the physical environment \cite{bajorunaite2023reality}, providing a lens to show parts of the physical environment \cite{wang2022realitylens}, or a radar-like widget that indicates the position of bystanders \cite{simeone2016vr}.
In contrast, one participant mentioned that system warnings, such as about a nearby wall, could be annoying, and users should be allowed to disable them. Keeping the user's safety in mind, the system could allow exceptions in certain locations.
Participants also expressed that not all places feel equally safe and that, as reported by Pavanatto et al. \cite{pavanatto2024working}, certain places are less suitable to use XR (Table \ref{tab:comparison-to-prior-work}).
They also mentioned that bystanders are usually curious, similar to what was reported by Pavanatto et al. \cite{pavanatto2024working}, but the initial curiosity wears off quickly.
However, the willingness to use an XR device in public seems to be highly individual, as was also reported by Cheng et al. \cite{cheng2025augmented} or Pavanatto et al. \cite{pavanatto2024working} (Table \ref{tab:comparison-to-prior-work}). While some participants reported doing this regularly, and one participant (P4) even used fully immersive environments in public, another (P2) refused to use it in public at all until its use becomes more widespread.
}

\newmarker{
{\textbf{XR should integrate physical keyboards.}}
All participants use physical keyboards when working with the AVP. Therefore, it is crucial to properly integrate keyboards in the virtual environment.
Also, ongoing research indicates that the typing performance on physical keyboards is still unmatched \cite{mcgill2015dose, knierim2018physical, grubert2018effects}.
AVP can track certain keyboards and make them visible in immersive environments.
As the supported keyboards might not be suitable for all users, a simple approach could be to have manual passthrough windows that users can place as needed, similar to what is possible in the immersed app.
}

\newmarker{
{\textbf{XR should allow flexible head fixations.}}
As reported in many prior studies~\cite{biener2022quantifying, biener2024hold, souchet2023narrative, cheng2025augmented}, the weight and comfort of head-mounted displays is still an issue (Table \ref{tab:comparison-to-prior-work}), specifically for long-term use~\cite{biener2022quantifying}.
The interviewed participants reported trying many different head-straps and the most comfortable head-strap seems to be highly subjective. Manufacturers should provide a variety of different solutions for their devices to accommodate the needs and preferences of a variety of users.
}

\newmarker{
{\textbf{XR-devices should be usable without a light-seal.}}
We also saw that participants liked head-straps which allowed them to use the device without a light seal to increase their field of view in the AR mode. This option is already available in optical see-through devices like the HoloLens but should also be considered by manufacturers in the video-see-through domain.
}

 \newmarker{
{\textbf{XR can potentially replace a variety of existing devices.}}
Participants reported using the AVP for a wide variety of tasks.
They also mentioned that, to some degree, it replaced several of their devices, including monitors and TV sets for viewing content on a larger scale, but also tablets as secondary displays or entertainment devices, and even smartphones for certain use cases.
The public sentiment analysis indicates that people tended to replace their screens more than their phones.
Yet, smartphones could certainly be replaced to an even higher degree, if the XR device would provide the necessary functionalities such as making calls.
}

\newmarker{
{\textbf{The resolution is good enough to work with virtual text.}}
Participants reported that the current resolution of the virtual display was good enough to also effectively work with text. 
}

\newmarker{
{\textbf{Passthrough resolution can limit certain use cases.}}
Even though the AVP provides one of the clearest passthrough experiences currently available, some use cases in which clear sight of the physical environment is necessary, such as when reading text on physical surfaces, are still limited.
}

\newmarker{
\subsection{Comparisons to Prior Research}
We found several similarities as well as differences between the findings from our interview study and prior research. Some of these were already mentioned in the previous subsections, and others are discussed in the following.
}

\newmarker{
In line with prior research that examined the use of large virtual displays~\cite{pavanatto2021we, pavanatto2024multiple, lisle2023spaces} participants mentioned the increased display space as one of the main advantages of using XR for work (Table \ref{tab:comparison-to-prior-work}).
Combined with reduced distractions in the XR environment it lead to a higher perceived productivity for most participants. 
This increase has also been indicated by Ruvimova et al.~\cite{ruvimova2020transport} who replaced an open office through a virtual environment, by Lee et al.~\cite{lee2022partitioning} who added virtual partitions around workspaces in AR, and recently for the AVP by Wienrich et al.~\cite{wienrich2025immersion} (Table \ref{tab:comparison-to-prior-work}).
}

\newmarker{
Participants used the AVP in various private and public places and reported that the virtual content arrangement was not affected by bystanders. Prior studies found that users preferred to keep displays within their proximity~\cite{ng2021passenger} and not to occlude others with the displays~\cite{medeiros2022shielding}.
Our participants were not concerned with this situation and prioritized consistency with their familiar working environment (Table \ref{tab:comparison-to-prior-work}). However, it is also possible that participants could not recall this behavior during the interviews, as they could be unconsciously taking the surrounding into account.
}

\newmarker{
Our participants indicated that, to some degree, they got used to wearing the device, as reported by Biener et al. \cite{biener2022quantifying, biener2024hold} (Table \ref{tab:comparison-to-prior-work}). For example, Biener et al.~\cite{biener2024hold} found that on the fifth day participants adjust and take off their devices less frequently as compared to the first day. 
Compared to the small virtual display used in the study by Biener et al.~\cite{biener2022quantifying}, our participants make full use of the XR capabilities to provide a very large display space. 
Multiple participants report using the device for many hours every day, indicating that the advantages already outweigh the drawbacks, such as the extra weight on the head (Table \ref{tab:comparison-to-prior-work}).
Some participants even use the device while eating or drinking, which was also observed by Biener et al. \cite{biener2024hold} during their one-week study, although the devices could affect eating frequency.
It seems that, if there are enough incentives or advantages of using XR, participants will accept the limitations.
Of course, participants could also be partially incentivized by the large amount of money that they paid for AVP.
}



\section{Limitations}

As our study focused on interviewing professionals who have used the device for an extended period of time, we cannot draw conclusions on changes in user behavior over time. Although we asked participants about any changes, the extended period of time since the first use of the device or the importance that participants may place on different aspects of their engagement with the system could have affected the responses. Long-term studies that evaluate user engagement and experience with XR remain necessary.
A major limitation of our study was that we were unable to collect objective information due to privacy considerations. As such, we could not collect views that support the participant's explanations or measure their engagement with the device during work.
Another limitation is the diverse background of the participants. While some used the device within the context of application development, others focused primarily on office tasks, resulting in varying benefits experienced by participants. We cannot conclude that XR is especially beneficial for a particular task.
Finally, we could not recruit participants who did not find the device suitable for their work and are no longer using it. Consequently, the responses we received from our participants might be biased to the positive side. In addition, we can only use feedback on initial experiences some of our participants shared, as well as public posts, to provide some insight into reasons why users choose not to use the device long-term.


\section{Conclusion}

In this paper, we present our findings from a series of interviews conducted with professionals who used the AVP for an extended period of time. 
The interviews highlighted that, as suggested by prior research, head-worn devices have the potential to increase the user's productivity and offer the benefit of a consistent, distraction-free workspace. 
At the same time, we identified issues that users face when using the AVP on a regular basis.
An important factor was the lack of professional tools that take advantage of the device's spatial tracking capabilities.
Therefore, most of the participants used AVP primarily to mirror their Mac workspace, rather than using native applications on the device for extended periods of time.
In addition, we want to highlight that it is important to make XR devices easily integrable into existing workflows and enable the user to customize them to optimally match their needs.
We encourage more long-term evaluations including objective measures to confirm productivity improvements. In addition, insights from users who opted out of the system after some use would provide a more differentiated view and highlight further areas that present road blocks for a more extensive adoption of XR in knowledge work.



\acknowledgments{
 This work was supported by 
 the Alexander von Humboldt Foundation funded by the German Federal Ministry of Education and Research,
 the German Research Foundation \emph{DFG} (grant 495135767),
 the Austrian Science Fund \emph{FWF} (grant I5912), and
 Snap Inc.
}


\bibliographystyle{abbrv-doi}
\balance
\bibliography{template}
\end{document}